\documentclass[fleqn,usenatbib]{mnras}

\usepackage{newtxtext,newtxmath}

\usepackage[T1]{fontenc}

\DeclareRobustCommand{\VAN}[3]{#2}
\let\VANthebibliography\thebibliography
\def\thebibliography{\DeclareRobustCommand{\VAN}[3]{##3}\VANthebibliography}

\usepackage{graphicx}	
\usepackage{amsmath}	
\usepackage{scalerel,stackengine}
\stackMath
\newcommand\reallywidehat[1]{%
\savestack{\tmpbox}{\stretchto{%
  \scaleto{%
    \scalerel*[\widthof{\ensuremath{#1}}]{\kern-.6pt\bigwedge\kern-.6pt}%
    {\rule[-\textheight/2]{1ex}{\textheight}}
  }{\textheight}%
}{0.5ex}}%
\stackon[1pt]{#1}{\tmpbox}%
}

\graphicspath{{./}{figures/}}

\title[Probing CGM with CHIME FRB]{A measurement of circumgalactic gas around nearby galaxies using fast radio bursts}

\author[X. H. Wu et al.]{
Xiaohan Wu,$^{1,2}$\thanks{E-mail: xwu@cita.utoronto.ca}
Matthew McQuinn,$^{3}$
\\
$^{1}$Canadian Institute for Theoretical Astrophysics, University of Toronto, Toronto, ON, Canada M5S 3H8\\
$^{2}$Harvard-Smithsonian Center for Astrophysics, 60 Garden Street, Cambridge 02138, MA, USA \\
$^{3}$Astronomy Department, University of Washington, Seattle, WA 98195, USA
}

\date{Accepted XXX. Received YYY; in original form ZZZ}

\pubyear{2015}

\begin{document}
\label{firstpage}
\pagerange{\pageref{firstpage}--\pageref{lastpage}}
\maketitle

\begin{abstract}
The distribution of gas in the circumgalactic medium (CGM) of galaxies of all types is poorly constrained.  Foreground CGMs contribute an extra amount to the dispersion measure (DM) of fast radio bursts (FRB).  We measure this DM excess for the CGMs of $10^{11}-10^{13}\ M_\odot$ halos using the CHIME/FRB first data release, a halo mass range that is challenging to probe in any other way.  Because of the uncertainty in the FRBs' angular coordinates, only for nearby galaxies is the localization sufficient to confidently associate them with intersecting any foreground halo.  Thus we stack on galaxies within $80$~Mpc, optimizing the stacking scheme to approximately minimize the stack's variance and marginalize over uncertainties in FRB locations. 
The sample has 20-30 FRBs intersecting halos with masses of $10^{11}-10^{12}\ M_\odot$ and also of $10^{12}-10^{13}\ M_\odot$, and these intersections allow a marginal $1-2\,\sigma$ detection of the DM excess in both mass bins.  The $10^{11}-10^{12}\ M_\odot$ halos bin also shows a DM excess at 1-2 virial radii. 
By comparing data with different models for the CGM gas profile, we find that all models are favored by the data up to 2-$\sigma$ level compared to the null hypothesis of no DM excess.
With 2000-3000 more bursts from a future CHIME data release, we project a 4-$\sigma$ detection of the CGM. 
Distinguishing between viable CGM models by stacking FRBs with CHIME-like localization would require tens of thousands of bursts.
\end{abstract}




\section{Introduction}
\label{sec:intro}

The circumgalactic medium (CGM) -- the diffuse gas that sits outside of galaxy disks and inside the halo virial radii -- is a crucial component in the baryonic processes in the universe.  It receives inflow of the intergalactic gas, fuels star formation in galaxies, and bears the impact of galactic feedback \citep{naab17, tumlinson17}.  An understanding of the CGM would resolve an aspect of the ``missing baryon problem'', that a substantial fraction of the baryons associated with halos has not been detected \citep{mcgaugh10, dai10, 2016ARA&A..54..313M}.

Questions remain about how the CGM is distributed around galaxies and how much mass there is in the CGM \citep[e.g.][]{tumlinson17}.  Fast radio bursts (FRB) are, however, starting to provide an unprecedented probe of the CGM with the advent of many radio telescopes geared for this science coming online across the globe \citep{cordes19, petroff19, petroff22, prochaska19, keating20}.  FRBs are bright (up to a few hundred Jy) transient radio pulses (typically up to a few milliseconds) of mostly extragalactic origin, and have been detected over a wide range of frequencies (400 MHz-8~GHz).  As the radio waves from FRBs travel through the intervening plasma, the interaction with free electrons causes a frequency-dependent delay of the arrival time, which is proportional to the integral of the electron number density along the line of sight -- the dispersion measure (DM).  Since most of the CGM gas is ionized, they contribute an extra amount to the total DM along a sightline, which can be used to constrain the CGM gas distribution \citep{mcquinn14, ravi19}.

There are many challenges to precisely measuring the excess DM from the CGM, especially in a sample of FRBs that are not well localized.  
First of all, the host galaxy and local environment around the FRB can contribute from a few tens to a few hundreds of pc~cm$^{-3}$ \citep{kulkarni15, connor16, tendulkar17, yang17, niu22, cordes22}.  Cosmic structures can also contribute scatter at the  hundreds of pc~cm$^{-3}$ level \citep{mcquinn14, macquart20}.   If the redshifts of the FRB host and intervening galaxies can be measured, requiring arcsecond localizations, one can subtract off the average cosmological contribution to the total DM, which significantly reduces the variance in the measurement.  Previous theoretical works predict that $\mathcal{O}(100)$ FRBs are required to put constraints on the CGM when stacking arcsecond-localized FRBs \citep{mcquinn14, ravi19}.  However, there are approximately 20 FRBs that are sufficiently well localized to date \citep{petroff22}, while thousands of predominantly unlocalized FRBs have been observed with more expected in the next few years.  
Without arcsecond localizations, many more FRBs are required to detect an excess DM from foreground CGMs \citep[but see][for constraining the Mikly Way DM using a small number of transients]{platts20}.  This paper considers this case.

We expand upon the recent work of \citet{connor21} and present a measurement of the CGM of nearby galaxies using the first FRB catalog published by the Canadian Hydrogen Intensity Mapping Experiment (CHIME) FRB project \citep{chimefrb}.  CHIME is a drift scan radio telescope operating across 400 MHz to 800 MHz.  The first CHIME/FRB catalog contains 535 FRBs detected between 2018 July 25 and 2019 July 1, including 18 repeating ones.  While the typical localization error of $0.2^\circ$ is too large for any sophisticated analysis on the CGM, nearby ($\lesssim100$~Mpc) halos that cover even larger areas on the sky ($>0.2^2$~deg$^2$) makes it possible to detect the DM excess from these halo CGMs.  We thus stack FRBs that intersect $10^{11}-10^{13}\ M_\odot$ halos within two virial radii and measure the DM excess, with a flexible weighting function that considerably reduces the stack's variance and, importantly, down-weights high-DM FRBs.  We show that with our weighting scheme, having 2000-3000 more FRBs from CHIME data release would lead to a $>3-4\,\sigma$ detection of the CGM of $10^{11}-10^{13}\ M_\odot$ halos, a number that is easily achievable with CHIME in the next few years.  This will open up a novel window for understanding the baryon physics in the universe, since other methods of studying the CGM such as the thermal and kinetic Sunyaev-Zeldovich effect \citep{schaan21} and halo X-ray emission \citep{chadayammuri22} usually probes higher mass ($>10^{12}-10^{13}\ M_\odot$) halos \citep[but see][for a measurement of nearby Milky Way size halos]{bregman21}.

This paper is organized as follows.  Section~\ref{sec:data} presents our galaxy catalog and FRB sample selection.  Section~\ref{sec:CHIME_DMexc} describes our weighting method and measurement of the DM excess from CGM.  Section~\ref{sec:models} compares our measurements with CGM models.  Section~\ref{sec:prediction} makes forecasts for future CHIME data release.

\section{Methods and results}
\label{sec:main}

\subsection{FRB selection}
\label{sec:data}

We use the non-repeaters in the CHIME FRB catalog.  Following \citet{connor21}, we exclude FRBs of Galactic latitudes within 5 degrees of the Galactic plane to avoid regions where the Milky Way DM is higher such that modeling errors for this contribution can be larger.  This gives a total number of 453 FRBs.  We have verified that our results are not sensitive to this Galactic latitude threshold.  For the FRB DM we use DM values provided by the CHIME catalog that have the Milky Way disk contribution removed using the \citet{ne2001} model.  We have checked that using the \citet{ymw16} model for the Milky Way disk DM does not change our conclusions.  We do not attempt to subtract the Milky Way halo DM since stacking should average out the fluctuations in the Milky Way halo DM, and the Milky Way halo DM is anticipated to be similar for all sightlines.  The Milky Way DM is also poorly constrained \citep{prochaska19, keating20, platts20}.

As in \citet{connor21}, we use the galaxies in the Gravitational Wave Galaxy Catalogue (GWGC) catalog \citep{gwgc}.  At 40~Mpc, the angular sizes of the virial radii of $10^{11}, 10^{12}, 10^{13}\ M_\odot$ halos are $0.15^\circ, 0.33^\circ, 0.70^\circ$ respectively, assuming $R_{\rm vir}=250(M/1.3\times10^{12}\ M_\odot)^{1/3}$~kpc.  We use this expression for $R_{\rm vir}$ throughout the paper.  Since the typical CHIME localization 1-$\sigma$ error is 0.2 degree, we use $10^{11}-10^{12}\ M_\odot$ halos at $0.5$-40~Mpc and $10^{12}-10^{13}\ M_\odot$ halos at $0.5$-80~Mpc for our analysis.  Later on, we develop a method to account for location uncertainties when comparing CGM models.
We calculate the halo masses of galaxies by converting from their stellar masses using the relation in \citet{moster10}.\footnote{Different models predict similar stellar mass-halo mass relations at $z=0$, e.g. Figure 34 of \citet{behroozi19}.}  To estimate the galaxy stellar mass, we adopt any of three methods depending on what photometric data is available (higher priority comes first): cross-matching with NASA-Sloan Atlas catalog\footnote{Provided by Matt Wilde.} (NSA, 4200 galaxies), JHK band SED fitting\footnote{We performed SED fitting using CIGALE \citep[][\url{https://cigale.lam.fr/}]{bouquien19} and find that using JHK bands alone provides more reliable stellar masses than using optical bands.  A lot of galaxies do not have optical photometry, as we queried the NASA/IPAC Extragalactic Database web service with astropy.} (3400 galaxies) or K-band mass-to-light ratio (100 galaxies), SDSS g-r color (400 galaxies).  For the remaining 1900 galaxies in GWGC for which we were unsuccessful at obtaining optical or infrared photometric data, we convert their B band luminosities to stellar masses by calibrating a conversion relation using the other galaxies.  Among the galaxies in the NSA catalog, 1800 also have JHK photometric data, for which we find that our stellar mass estimates using JHK bands mostly lie within a factor of 3 from the NSA stellar masses.  We have verified that our main conclusions do not change if we use the JHK stellar masses for these galaxies instead.

For the candidate $>10^{11}\ M_\odot$ halos, we identify galaxy groups and remove satellite galaxies.  For a given galaxy, we define it as a satellite if it lies within $1.2\ R_{\rm vir}$ of a more massive galaxy nearby in terms of their 3D distance.  We do not consider M33 as a satellite of M31.  Since the distances listed in GWGC have a typical error of 20\%, we also test determining a satellite using the 2D projected distance and the difference in the radial velocities.  We find that our results change very little if a satellite is identified by requiring that the radial velocities differ within 3 times the halo circular velocity instead of using the 3D distances.  Our group finding leaves 4600 $10^{11}-10^{12}\ M_\odot$ galaxies at $0.5$-40 Mpc and 4000 $10^{12}-10^{13}\ M_\odot$ galaxies at $0.5$-80 Mpc.
We also remove 5 FRBs that we identify to intersect $>10^{13}\ M_\odot$ halos at $<1\ R_{\rm vir}$.

\begin{figure*}
\includegraphics[width=2\columnwidth]{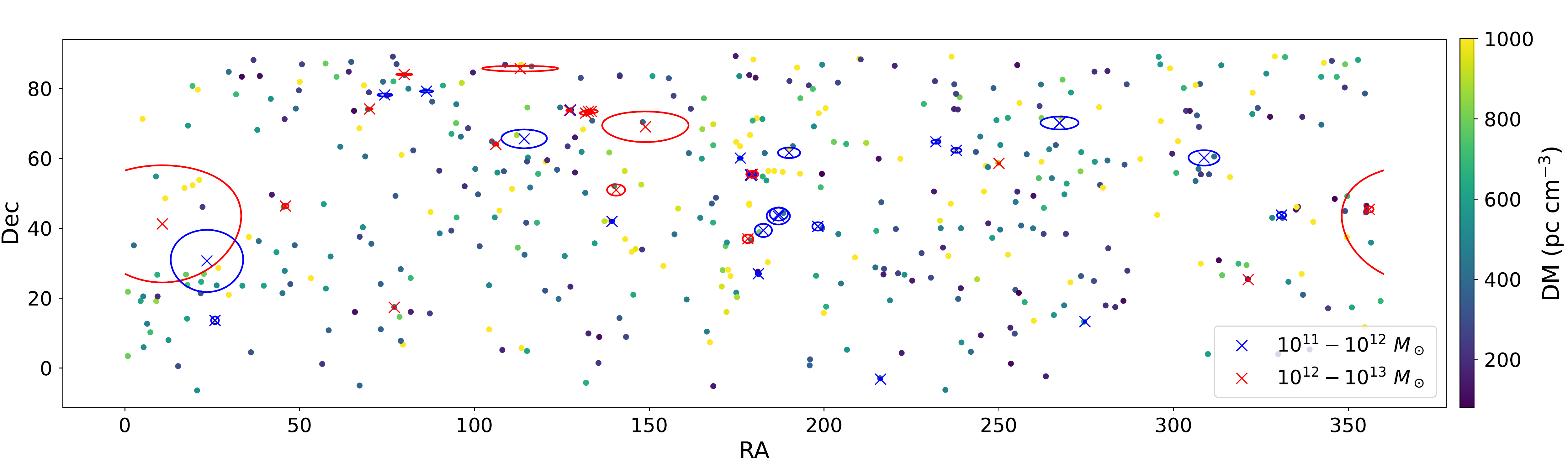}
\caption{Locations of the CHIME catalog FRBs used in our analysis as well as of galaxies having FRB intersection with an impact parameter of $b<1\ R_{\rm vir}$.  Dots show the position of FRBs, with colors denoting their DM.  Blue and red crosses illustrate the centers of $10^{11}-10^{12}\ M_\odot$ and $10^{12}-10^{13}\ M_\odot$ halos, respectively.  Ellipses around these halos map a circle of radius $1 R_{\rm vir}$ into these angular coordiantes.}
\label{fig:vis_1Rvir}
\end{figure*}

\begin{table*}
\centering
\caption{Number of FRBs and halos in each impact parameter ($b$) bin and the measured DM excess in units of parsec~cm$^{-3}$..  From top to bottom: $b\in[0,1], [1,1.5], [1.5,2]\ R_{\rm vir}$.  {\it Second to fourth column:} number of FRBs intersecting $10^{11}-10^{12}\ M_\odot$ halos, number of FRBs intersecting M33, and the number of $10^{11}-10^{12}\ M_\odot$ halos with FRB intersections.  Numbers in parentheses indicate the 1-$\sigma$ range (16$^{\rm th}$ and 84$^{\rm th}$ percentiles) of the number of FRBs owing to localization uncertainties, calculated by perturbing FRB locations assuming Gaussian errors.  {\it Fifth column:} the measured DM excess and 1-$\sigma$ error, with numbers in brackets denoting the results excluding M33.  {\it Sixth to ninth column:} the same but for $10^{12}-10^{13}\ M_\odot$ halos.}
\label{tab:n_frbs}
\begin{tabular}{l|cccc|cccc}
\hline
$b$ range & $N_{\rm FRB}$ & $N_{\rm FRB}$ & $N_{\rm halo}$ & DM excess & $N_{\rm FRB}$ & $N_{\rm FRB}$ & $N_{\rm halo}$ & DM excess \\
& $10^{11}-10^{12}$ & M33 & $10^{11}-10^{12}$ & & $10^{12}-10^{13}$ & M31 & $10^{12}-10^{13}$ & \\
\hline
$[0,1]\ R_{\rm vir}$ & 26 (26-31) & 6 & 24 & $71\pm53 (18\pm61)$ & 30 (28-33) & 16 & 18 & $69\pm49 (96\pm72)$ \\
$[1,1.5]\ R_{\rm vir}$ & 31 (28-34) & 5 & 26 & $87\pm48 (78\pm53)$ & 38 (36-42) & 22 & 18 & $-4\pm43 (6\pm61)$ \\
$[1.5,2] R_{\rm vir}$ & 39 (32-40) & 9 & 32 & $41\pm43 (35\pm46)$ & 57 (49-56) & 28 & 31 & $37\pm35 (-11\pm46)$ \\
\hline
\end{tabular}
\end{table*}

We select FRBs that intersect nearby halos out to $2\ R_{\rm vir}$ and bin then according to their $b/R_{\rm vir}$.  Figure~\ref{fig:vis_1Rvir} shows the map of FRBs, with colors indicating the DM values, and locations of galaxies that have FRB intersections within $1\ R_{\rm vir}$.  Blue and red crosses represent the centers of $10^{11}-10^{12}\ M_\odot$ and $10^{12}-10^{13}\ M_\odot$ halos in this sample, respectively.  Ellipses of corresponding colors map a circle of radius $1\ R_{\rm vir}$ into RA-Dec coordinates.  The two biggest circles at RA $<50$~deg are M31 and M33.  With our selection choices, there are 26 FRBs intersecting 24 halos with masses of $10^{11}-10^{12}\ M_\odot$ within $1\ R_{\rm vir}$, and 30 FRBs intersecting 18 halos with $10^{12}-10^{13}\ M_\odot$, although half of these intersections are with M31.  Among the 53 FRBs selected, 6 pass through two halos.  Table~\ref{tab:n_frbs} lists the number of FRBs and halos falling in impact parameter bins of $[0,1], [1,1.5], [1.5,2]\ R_{\rm vir}$.  The numbers in brackets in this table represent the 1-$\sigma$ error on these numbers owing to localization uncertainties, obtained by perturbing FRB locations assuming Gaussian errors.\footnote{We note that the error-bars on the number of FRBs in a bin are likely asymmetric around the number calculated using the maximum likelihood localization.  This occurs because some FRBs fall on the edge of halos.}  If a FRB falls into multiple radial bins owing to intersections with different halos, we only count this FRB once, grouping it with the halo it intersects at the smallest impact parameter.  We have tested that counting the halo twice does not change our results.

\subsection{Measuring DM excess}
\label{sec:CHIME_DMexc}

\begin{figure}
\includegraphics[width=\columnwidth]{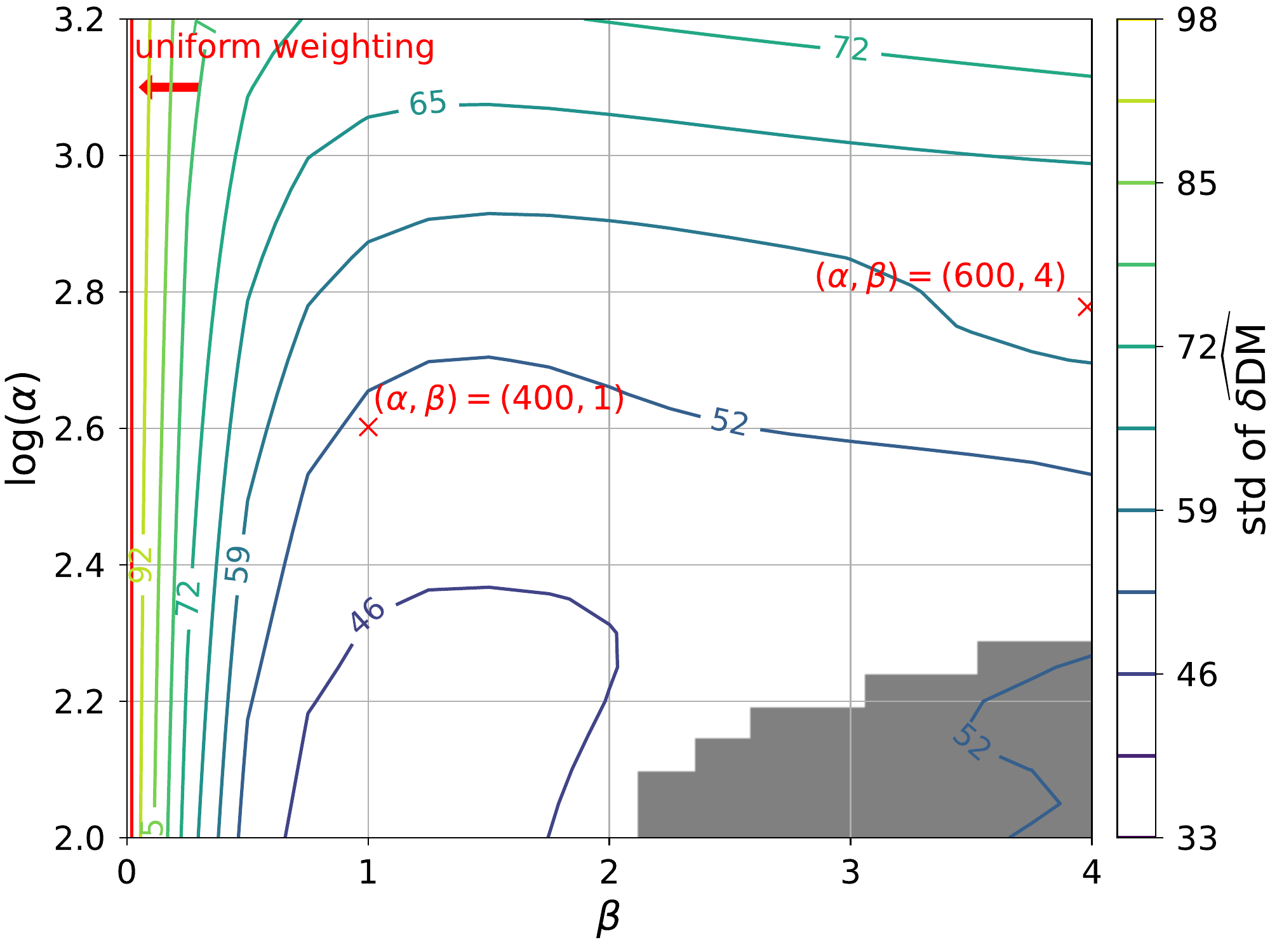}
\includegraphics[width=\columnwidth]{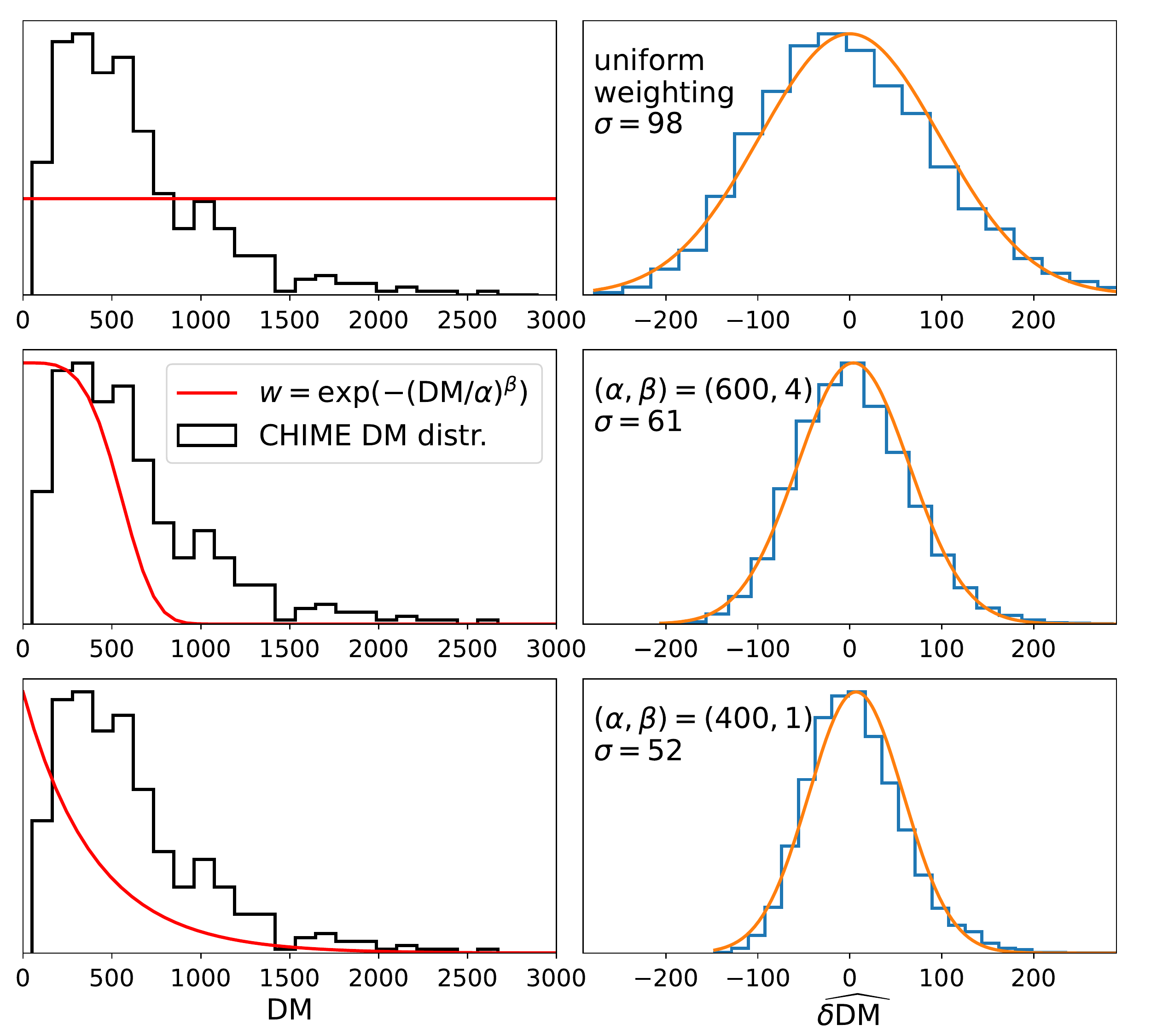}
\caption{Top panel: contours of the standard deviation (std) of the weighted-average DM as a function of the parameters of our weight function $\alpha$ and $\beta$ (c.f. eqn~\ref{eqn:weights}), obtained by randomly sampling 20 FRBs from the CHIME DM distribution.  The gray shaded region represents $\alpha,\beta$ values that can suppress all sampled DM values.  The line at $\beta=0$ and red crosses indicate the weighting functions we examine in the bottom panels: uniform weighting, $(\alpha, \beta) = (600, 4)$, $(\alpha, \beta) = (400, 1)$.  Left column of the following rows: the black lines show the DM histogram of the CHIME FRBs, and red lines represent the corresponding weighing functions in arbitrary units.  Right column: histogram of the excess weighted-average DM (equation~\ref{eq:DM_estimator}), for different values of $\alpha, \beta$, and uniform weighting (blue lines).  Orange lines represent Gaussian fits to the histograms.  Also quoted are the standard deviations of the weighted-mean DM. All DM units are in parsec~cm$^{-3}$.
\label{fig:DMstd}}
\end{figure}

We aim to examine whether the FRBs with impact parameter $b<2\ R_{\rm vir}$ exhibit a statistically significant excess DM over the full sample of FRBs.
A simple estimator for the DM excess is the sample mean DM minus the mean of all CHIME FRBs.  However, since the number of FRBs with halo intersections is small, one or two FRBs with high DM can significantly affect the sample mean DM.  For a sample size of 20, one FRB with DM~$=2000$~pc~cm$^{-3}$ contributes 100~pc~cm$^{-3}$ to the sample mean, and the CHIME FRBs have a high-DM tail of DM~$\gtrsim1500$ (see the black histogram in Figure~\ref{fig:DMstd}; hereafter we omit the unit of DM pc~cm$^{-3}$).
An unweighted mean is thus not an ideal estimator for the DM excess as high DM bursts add to the variance without contributing much signal.  Therefore, instead of calculating the sample mean DM, we compute a weighted-average DM where we down-weight contributions from high-DM FRBs using a flexible weighting function
\begin{equation}
w({\rm DM}) \propto \exp\left( -({\rm DM}/\alpha)^\beta \right),
\label{eqn:weights}
\end{equation}
where $\alpha$ and $\beta$ shape how quickly the weight cuts off.
If a sub-sample of FRBs ($\mathcal{I}$) have a constant DM excess $\delta{\rm DM}$, an estimate for $\delta{\rm DM}$ is
\begin{equation}
\widetilde{\delta{\rm DM}} = \frac{1}{\sum_{i\in\mathcal{I}} w({\rm DM}_i)} \sum_{i\in\mathcal{I}} w({\rm DM}_i) ({\rm DM}_i) - \overline{{\rm DM}},
\end{equation}
where $i$ denotes the indices of FRBs and
\begin{equation}
\overline{{\rm DM}} = \frac{1}{\sum_i w({\rm DM}_i)} \sum_i w({\rm DM}_i){\rm DM}_i
\end{equation}
is the weighted mean DM of the whole CHIME sample.
This down-weighting is a major difference of our work from \citet{connor21}, who used uniform weighting.  Optimizing our weighting function allows us to detect a DM excess with added precision.

One worry with such weighting is that sightlines with excess DM will receive smaller weights, such that the weighted average will return a smaller excess.  Specifically, since our weighting function is a non-linear function of the DM excess, if we input an array of DM values drawn from some the CHIME DM distribution and artificially put in an excess $\delta$DM, the estimator becomes
\begin{align*}
\widetilde{\delta{\rm DM}} &= \frac{1}{\sum_i w({\rm DM}_i)+\delta{\rm DM}} \sum_i w({\rm DM}_i+\delta{\rm DM})({\rm DM}_i+\delta{\rm DM})& \\
&-\; \frac{1}{\sum_i w({\rm DM}_i)} \sum_i w({\rm DM}_i){\rm DM}_i.
\end{align*}
This does not return $\delta{\rm DM}$ but rather is biased by a factor that we denote by
\begin{equation}
f = \widetilde{\delta{\rm DM}} / \delta {\rm DM},
\end{equation}
which conveniently we find is nearly independent of the DM excess for $\delta{\rm DM}<200$, a limit that is easily satisfied in models for the foreground CGMs of galactic halos.  This independence of $f$ is convenient as it means that the correction does not depend on the true excess. 
For a sub-sample $\mathcal{I}$ that has a {\it constant} DM excess $\delta$DM, an estimator for $\delta$DM is given by iteratively solving
\begin{equation}
\reallywidehat{\delta{\rm DM}} = \frac{1}{\sum_{i\in\mathcal{I}} w\left({\rm DM}_i - \reallywidehat{\delta{\rm DM}}\right)} \sum_{i\in\mathcal{I}} w\left({\rm DM}_i - \reallywidehat{\delta{\rm DM}}\right)\ {\rm DM}_i - \overline{{\rm DM}}.
\label{eq:calc_DMexc}
\end{equation}
This iterative estimator takes out the bias by removing the weightings' dependence on $\delta DM$. For our DM excess of interest ($\lesssim100$), the solution roughly converges at 
\begin{equation}
\reallywidehat{\delta{\rm DM}} \approx \widetilde{\delta{\rm DM}} / f.
\label{eq:DM_estimator}
\end{equation}
In the limit of a large sample, the above equation gives the unbiased excess, $\delta{\rm DM}$.
We use this estimator, $\reallywidehat{\delta{\rm DM}}$, to measure the DM excess of FRBs with CGM intersections.  We explore the best choices for $\alpha, \beta$ in what follows.

We examine how much reduction in the variance of the weighted-average DM that our weighting function can give.  To this end, we created a mock FRB catalogue of size $10^5$ that has the same DM distribution as the CHIME catalog.  We then randomly sample $N_{\rm DM}$ FRBs from this mock catalogue and for each sample calculate $\reallywidehat{\delta{\rm DM}}$ using equation~(\ref{eq:DM_estimator}) for a range of $\alpha, \beta$ values.  We repeat this step 10,000 times and compute the standard deviation (std) of $\reallywidehat{\delta{\rm DM}}$ for each $\alpha, \beta$.

The top panel of Figure~\ref{fig:DMstd} shows contours of the std of $\reallywidehat{\delta{\rm DM}}$ as a function of $\alpha, \beta$, obtained by randomly sampling $N_{\rm DM} =20$ FRBs from the CHIME DM distribution.  Twenty is chosen to roughly match the number of CHIME FRBs with halo intersections, although we find that the contours of the std of $\reallywidehat{\delta{\rm DM}}$ remain mostly unchanged when assuming different $N_{\rm DM}$ such as $N_{\rm DM}=100$.  The gray shaded region represents $\alpha,\beta$ values that can lead to all zero weights to machine precision, for at least one time out of the 10,000 times of random sampling.  The best estimator for mean DM when sampling 20 FRBs seems to be given by values near $(\alpha, \beta) = (100, 1)$, but such a weighting falls off exponentially above the low value of ${\rm DM}\sim100$, selecting the few sightlines with the smallest DM.  Therefore, the minimum variance estimator is close to selecting the minimum DM -- for many distributions the minimum value in the DM array converge faster than the mean.  Such an estimator is problematic because the minimum is more affected by problems such as some FRBs being at very low-redshift in the foreground of our halos \citep[e.g. the low-DM FRBs in][]{bhardwaj21a, bhardwaj21b}.  We investigate this possibility in Appendix~\ref{sec:discussion}, finding that it is likely that $\sim 5$ of our FRBs out of 20-30 are likely to originate from the galaxy in the halo that they are identified to intersect or from a foreground galaxy; this contamination is problematic for the minimum variance estimator.  We thus explore $(\alpha,\beta) = (600,4)$ and $(400,1)$, which -- while giving substantial weight to about half of the FRBs -- still give a factor of $1.6$ and 2 reduction of the variance relative to an unweighted estimator, respectively, and just modestly larger variances than the minimum variance $(\alpha, \beta) = (100, 1)$.\footnote{The bias factor $f$ of the estimator with $(\alpha,\beta) = (600,4)$ is $0.65$, while $(400,1)$ is unbiased.}

The black lines in the left column of Figure~\ref{fig:DMstd} illustrate the DM histogram of the CHIME FRBs, and red lines represent these weighing functions in arbitrary units.  From top to bottom we use uniform weighting, $(\alpha, \beta) = (600, 4)$, and $(\alpha, \beta) = (400, 1)$.
Using $(\alpha, \beta) = (600, 4)$ has a much stronger cut-off at high DM values, leading to 20-30\% of the CHIME DM distribution contributing negligibly.  The right panels of Figure~\ref{fig:DMstd} show the histograms of $\reallywidehat{\delta{\rm DM}}$ by randomly sampling 20 FRBs from mocks.  Blue lines illustrate the histograms, and the orange lines represent Gaussian fits to the histograms.  Also quoted are the standard deviations of $\reallywidehat{\delta{\rm DM}}$.
Since $(\alpha, \beta) = (400, 1)$ gives a slightly more skewed distribution of the weighted-mean DM with a high-value tail, by default we use $(\alpha, \beta) = (600, 4)$, but we have verified that $(\alpha, \beta) = (400, 1)$ yields very similar conclusions.

\begin{figure*}
\includegraphics[width=2\columnwidth]{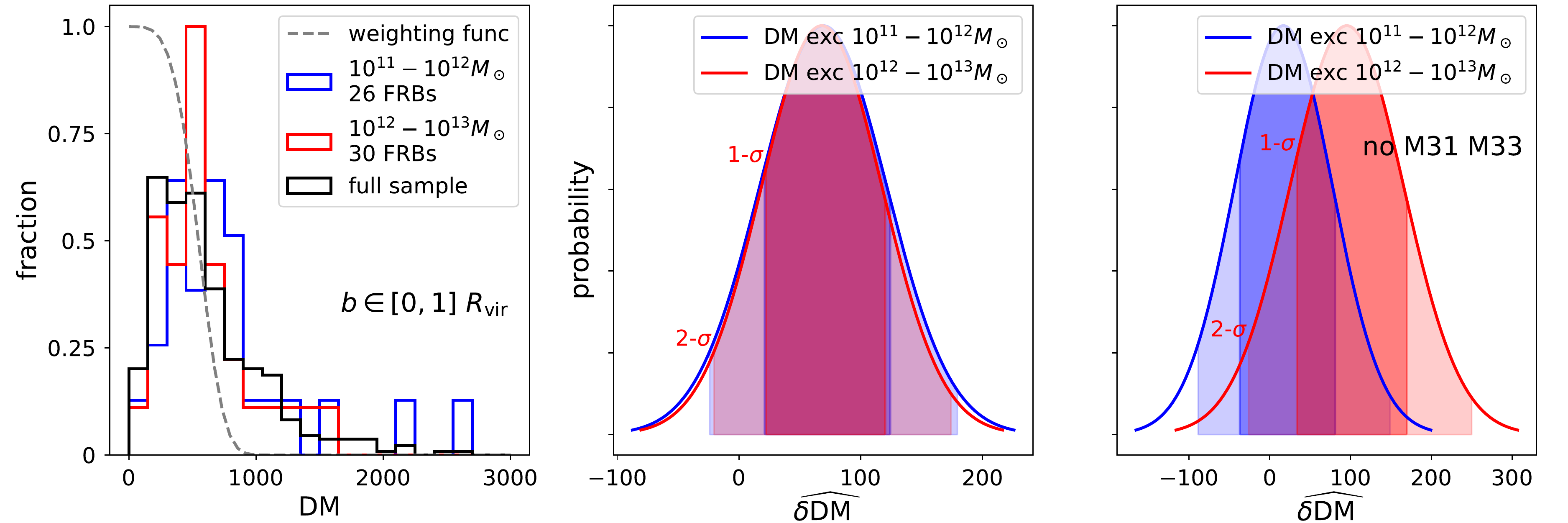}
\caption{Statistics of CHIME FRBs that intersect nearby $10^{11}-10^{12}\ M_\odot$ (blue) and $10^{12}-10^{13}\ M_\odot$ (red) halos within $1\ R_{\rm vir}$.
{\it Left panel:} the blue and red histograms show the DM distributions of the 26 and 30 FRBs intersecting the smaller and larger halo mass bins, respectively.  The black line illustrates the DM distribution of the full CHIME sample.  The gray dashed line represents our weighting function in arbitrary units, which to calculate the stacked DM is effectively multiplied by the DM distribution.
{\it Middle panel:} The blue and red lines represent the likelihood of our estimator $\widehat{\delta{\rm DM}}$ with $(\alpha, \beta) = (600, 4)$ for the whole CHIME sample, where the central value is our measurement -- the estimator applied to the observational data -- and the Gaussian PDF is calculated by random sampling of the CHIME DM distribution with 26 and 30 FRBs respectively.   
Shaded regions show the 1-$\sigma$ and 2-$\sigma$ bounds of the distributions.
{\it Right panel:}  The same as the middle panel, except that we have excluded FRBs that intersect M31 and M33, which results in 20 FRBs intersecting $10^{11}-10^{12}\ M_\odot$ halos and 15 intersecting $10^{12}-10^{13}\ M_\odot$ ones.
\label{fig:observed_hists_and_vals}}
\end{figure*}

Using the FRBs that intersect halos, we measure their excess DM by using the estimator given by equation~(\ref{eq:DM_estimator}).
The left panel of Figure~\ref{fig:observed_hists_and_vals} shows the DM distributions of the 26 FRBs that intersect $10^{11}-10^{12}\ M_\odot$ halos (blue) within $1\ R_{\rm vir}$, and the 30 FRBs intersecting $10^{12}-10^{13}\ M_\odot$ halos (red).  The black line represents the DM distribution of the whole CHIME sample.  The gray dashed line represents our weighting function.
The middle panel shows the probability distributions of the measured DM excess, centered on the solution of equation~(\ref{eq:DM_estimator}) with the std determined by randomly sampling 26 (blue line) and 30 (red line) FRBs from the CHIME DM distribution respectively.  The shaded regions indicate the 1-$\sigma$ and 2-$\sigma$ bounds of the distributions.  The excess DM of both halo groups seems to be detected at $1.5$-$\sigma$ level.
Owing to the large extent of M31 and M33 on the sky, we also performed our analysis without the FRBs intersecting these galaxies.  This gives 20 FRBs intersecting $10^{11}-10^{12}\ M_\odot$ halos and 14 intersecting $10^{12}-10^{13}\ M_\odot$ ones.  The right panel shows the distributions of the excess DM when excluding M31 and M33.  While the DM excess from $10^{12}-10^{13}\ M_\odot$ halos can still be detected at $1.3$-$\sigma$ level, an insignificant excess in the $10^{11}-10^{12}\ M_\odot$ mass bin is statistically preferred.

Our 1-$1.5$\;$\sigma$ level detection of the foreground CGM is less significant than the $>2$ $\sigma$ level of detection in \citet{connor21}, even though the std of our estimator is a factor of two smaller.  However, there are two FRBs with ${\rm DM}>2000$ intersecting $10^{11}-10^{12}\ M_\odot$ halos (left panel of Figure~\ref{fig:observed_hists_and_vals}).  Without a weighting function, these FRBs raise the sample mean DM by over 100, for a sample size of 30.  This likely leads to the t-test returning $p<0.05$ in \citet{connor21}.

\begin{figure*}
\includegraphics[width=2\columnwidth]{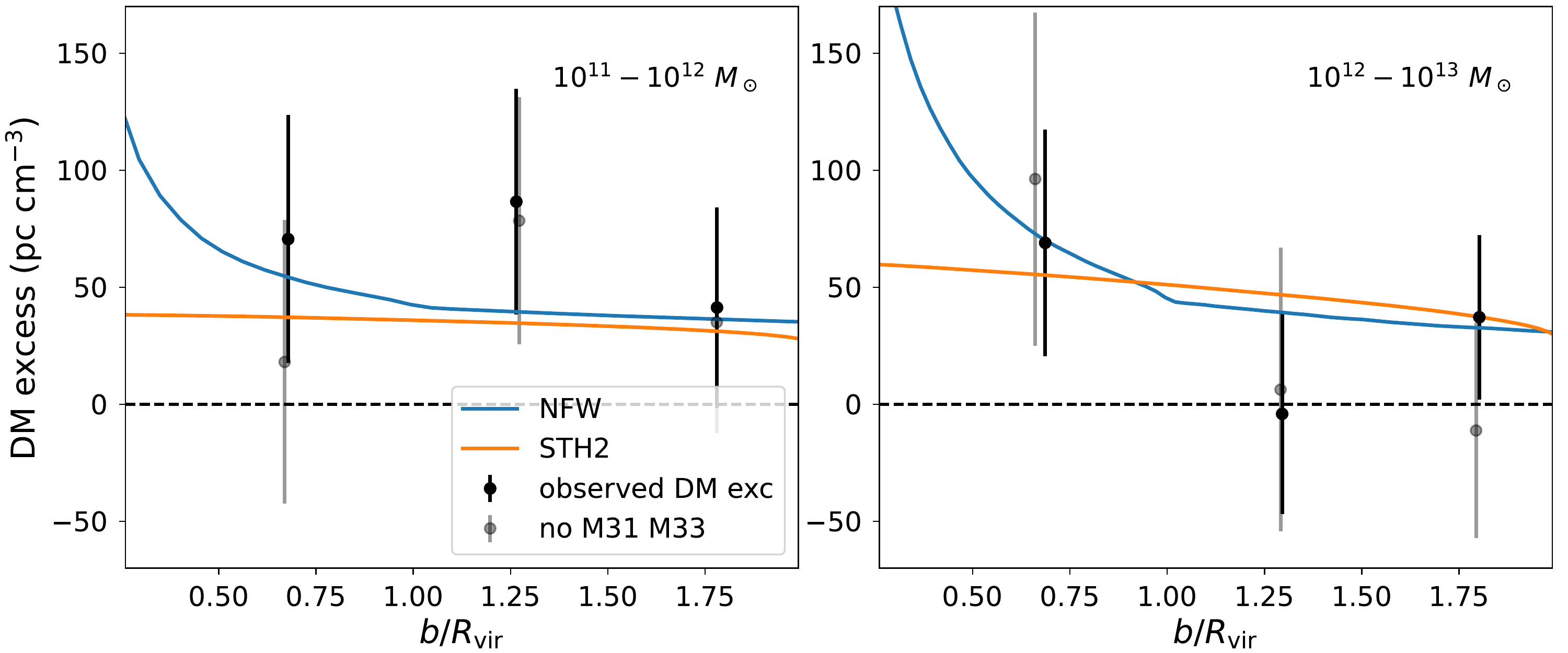}
\caption{The measured and model-predicted DM excess as a function of impact parameter at different impact parameters $b$ to the foreground halo.  Left and right panels show our results for the $10^{11}-10^{12}\ M_\odot$ and $10^{12}-10^{13}\ M_\odot$ halos, respectively.  Black points with errorbars illustrate the observed DM excess including all halos, and the gray points excluding M33 (left panel) or M31 (right panel).  As a reference for the expected signal amplitude, the blue and orange lines show the radial DM profiles in the NFW-tracing and the $2\,R_{\rm vir}$ spherical top hat (STH2) models for the CGM gas, respectively (see \S~\ref{sec:models}). These models are calculated assuming halo masses of $4\times10^{11}\ M_\odot$ (left) and $4\times10^{12}\ M_\odot$ (right).}
\label{fig:DMexc_radial_profile}
\end{figure*}

We performed the same analysis on the DM excess of FRBs intersecting halos within impact parameters of $1 - 1.5\ R_{\rm vir}$ and $1.5 -2\ R_{\rm vir}$.  We find that the FRBs intersecting $10^{11}-10^{12}\ M_\odot$ halos continue to show a DM excess at $\gtrsim1$-$\sigma$ level in these bins, while the $10^{12}-10^{13}\ M_\odot$ halos no longer get a detection of DM excess.
Figure~\ref{fig:DMexc_radial_profile} summarizes our results.  Dots with error-bars show the measured DM excess as a function of $b/R_{\rm vir}$ assuming constant DM excess in each radial bin.  Left and right panels illustrate results for the $10^{11}-10^{12}\ M_\odot$ and $10^{12}-10^{13}\ M_\odot$ halos, respectively.  The black and gray colors indicate including all halos and excluding M31 and M33, respectively.  The $b$ values are calculated as the weighted-average $b$ of all FRBs, with the weights given by our weighting function and the DM.  As a reference for the expected DM excess, this figure also shows are the radial DM profiles for a model in which the CGM gas traces the NFW profile and another where it is distributed as a top hat with radius of $2\,R_{\rm vir}$ (STH2) model.
Our DM excess measurements are largely consistent with the model predictions.
Both models are discussed in what follows.

\subsection{Comparison with CGM models}
\label{sec:models}

Since our constraints are in the ballpark of the expected CGM DM excess, it is possible that they already rule out some CGM models.
Suppose we are in the limit of a very large number of FRBs, although with imprecise localization.  We expect the following relation to hold for the sub-sample ($\mathcal{I}$) of FRBs that have CGM intersections:
\begin{equation}
\frac{1}{\sum_i w_{i\in\mathcal{I}}} \sum_{i\in\mathcal{I}} \left[{\rm DM}_i - \delta {\rm DM}_i(b, M)\right] w_i = \overline{{\rm DM}},
\end{equation}
where $i$ is the index of a FRB, ${\rm DM}$ is its measured DM, $\delta {\rm DM}_i$ is its excess DM from foreground CGMs, and $w_i$ is the weight calculated from the no-excess DM (i.e. from ${\rm DM}_i - \delta {\rm DM}_i$).  $\overline{{\rm DM}}$ is the weighted-average DM of the whole CHIME sample.
We note that unlike equation~(\ref{eq:calc_DMexc}), $\delta {\rm DM}_i(b, M)$ models the excess DM in terms of the impact parameter $b$ and halo mass $M$ of each foreground galaxy (and not necessarily just one halo despite our notation).

We aim to find the model for $\delta {\rm DM}_i$ that best describes the data.
We therefore define a measure for the goodness of fit
\begin{equation}
\xi^2 = \left( \left\{ \frac{1}{\sum_{i\in\mathcal{I}} w_i} \sum_{i\in\mathcal{I}} \left[{\rm DM}_i - \delta {\rm DM}_i (b, M) \right] w_i \right \} - \overline{{\rm DM}} \right)^2.
\label{eq:chi2}
\end{equation}
We can also take the expectation value of $\xi^2$ by random sampling the same number of intersections from the full DM distribution:
\begin{equation}
\langle\xi^2\rangle = \Bigg\langle \left( \frac{1}{\sum_i w({\rm DM}_i)} \sum_i w({\rm DM}_i) {\rm DM}_i - \overline{{\rm DM}} \right)^2 \Bigg\rangle,
\end{equation}
where the summation goes over the desired number of FRBs $N_{\rm DM}$.
The expectation scales with $N_{\rm DM}$ as $\langle\xi^2\rangle \propto N_{\rm DM}^{-1}$.

Our formalism still holds when dividing the FRBs into impact parameter bins based on their $b/R_{\rm vir}$ values, as long as there are enough FRBs in a bin.  For each radial bin $j$, we can calculate the $\chi^2$ statistic by dividing $\xi^2$ by $\langle \xi^2 \rangle$ in that bin:
\begin{equation}
\chi^2_j = \frac{\xi_j^2}{\langle\xi^2\rangle_j} \left( \mathbf{b} \right),
\end{equation}
which is a function of the set of impact parameters $\mathbf{b}$.
Summing these over radial bins gives the total:
\begin{equation}
\chi^2 = \sum_j \chi^2_j.
\end{equation}

The above is the procedure without accounting for localization error.  However, for some of our intersections, the localization is comparable to the halo viral radius. To take into account of the localization uncertainties, we calculate the marginalized $\chi^2$ value of bin $j$:
\begin{equation}
\exp\left(-\chi^2_j/2\right) = \int \exp\left(-\frac{1}{2}\frac{\xi^2_j}{\langle\xi^2\rangle_j}\left(\mathbf{b}\right)\right)\ P(\mathbf{b}) \ \mathrm{d}\mathbf{b}.
\label{eq:marg_chi2_bin}
\end{equation}
The total $\chi^2$ is
\begin{equation}
\exp\left(-\chi^2/2\right) = \int \exp\left(-\frac{1}{2}\sum_j\frac{\xi^2_j}{\langle\xi^2\rangle_j}\left(\mathbf{b}\right)\right)\ P(\mathbf{b})\ \mathrm{d}\mathbf{b}.
\label{eq:marg_chi2_tot}
\end{equation}

Given a model of the CGM, the radial profile of DM around a sample of halos has a one-halo term owing to the gas surrounding these halos themselves, and also a rather flat two-halo term coming from the overlapping of gas from other halos.  We use the DM radial profiles calculated using the {\sf CGMBrush} algorithm \citep{williams22} for our $\chi^2$ calculation.  This algorithm subtracts off the dark matter associated with each halo in an N-body simulation and then pastes on different models for the distribution of the gas associated with that halo.  It assumes that gas outside of halos traces the dark matter.\footnote{To compute the total DM excess for a FRB for the few sightlines that have multiple intersections, we first sum up the one-halo terms from all the nearby $10^{11}-10^{13}\ M_\odot$ halos that this FRB can intersect along its sightline.  We compute an average two-halo term using these halos and add to the total excess DM, since an FRB may intersect multiple halos at $>1\ R_{\rm vir}$. It perhaps makes the most sense to take the maximum value of the two-halo terms if the intersecting halos are within several impact parameters, and to sum the two halo the two-halo terms if halos are further apart than several impact parameters.  However, the exact algorithm does not affect our results by $\Delta \chi^2 >1$.}
Using these models for the DM excess, we calculate $\chi^2$ values using equations~(\ref{eq:chi2}) and (\ref{eq:marg_chi2_bin}) for FRBs in each of the three radial bins $[0,1], [1,1.5], [1.5,2]$, in units of $R_{\rm vir}$.  To speed up the calculation, we restrict to using FRBs that lie within $(2\ R_{\rm vir} + 3\times0.2^\circ)$ of galaxies, where $0.2^\circ$ is the typical localization error of CHIME.  As mentioned in Section~\ref{sec:data}, we count each FRB only once if any intersects multiple halos.  Counting galaxy-FRB pairs instead leads to somewhat larger $\Delta\chi^2$, but with differences smaller than 1-2.  To evaluate the high-dimension integral on the right hand side of equation~(\ref{eq:marg_chi2_bin}) or equation~(\ref{eq:marg_chi2_tot}), we draw from a 2D Gaussian for the location of each FRB on the sky plane so that the draws trace $ P(\mathbf{b})$ and sum up $10^4$ realizations for a Monte Carlo evaluation.

We perform this $\chi^2$ calculation for the $2\, R_{\rm vir}$ spherical top hat (STH2) model of the CGM and the NFW model.  These extreme models roughly bound the DM excess predicted by the more realistic models in \citet{williams22}.  The STH2 model has only an eighth of the halo-associated baryons within $1\,R_{\rm vir}$, whereas the NFW contains all of them  within $1\,R_{\rm vir}$.  Indeed, owing to its diffuseness, we find that the STH2 DM excess is shaped at most radii by the two-halo term and not the halo profile.   Figure~\ref{fig:DMexc_radial_profile} illustrates DM excess as a function of $b/R_{\rm vir}$ predicted by the NFW (blue lines) and the STH2 (orange lines) models.  These illustrative curves in the left and right panels assume halo masses of $4\times10^{11}\ M_\odot$ and $4\times10^{12}\ M_\odot$ respectively, roughly the mean of our two halo samples.  Despite these two profile making much different assumptions for the halo associated baryons, 
these models give similar predictions outside $0.5\ R_{\rm vir}$, with DM differences smaller than 50 because of the large two-halo excess that depends weakly on the halo gas profiles.  Our sample contains only 7 and 4 FRBs intersecting with $b<0.5\ R_{\rm vir}$ the small and large halo mass bins respectively.  The STH2 model predicts a flat DM excess of 30-50 at $0.5-1\ R_{\rm vir}$ for the $10^{11}-10^{12}\ M_\odot$ halos, and 50-80 for the $10^{12}-10^{13}\ M_\odot$ halos.  The NFW model produces 30-50 higher DM excess.  The two-halo term is about 30-50 for both halo mass bins outside $1\ R_{\rm vir}$.  Because the DM differences among models are only $\gtrsim50$, we expect the $\chi^2$ values of these two models (and other models in \citealt{williams22}) to differ only at $<1$-$\sigma$ level.

\begin{table*}
\centering
\caption{$\chi^2$ and $\Delta\chi^2$ values of different models in each impact parameter ($b$) bin, for the $10^{11}-10^{12}\ M_\odot$ (upper) and $10^{11}-10^{12}\ M_\odot$ (lower) halos.  From top to bottom: $b\in[0,1], [1,1.5], [1.5,2]\ R_{\rm vir}$, and the total $\chi^2$ considering all bins.  For each $\chi^2$ value we list the $\chi^2$ values by marginalizing over localization uncertainties, and the number in brackets shows the $\chi^2$ obtained by using the maximum likelihood localization.  Second column: $\chi^2$ of the no DM excess model.  Third and fourth columns: $\Delta\chi^2$ of no DM excess minus the $2R_{\rm vir}$ spherical top hat model (STH2), and the NFW model.  Fifth column: $\Delta\chi^2$ of no DM excess minus the NFW model, excluding M33 or M31.}
\label{tab:chi2}
\begin{tabular}{lcccc}
\hline
$10^{11}-10^{12}\ M_\odot$ & $\chi^2$ & $\Delta\chi^2$ & $\Delta\chi^2$ & $\Delta\chi^2$ no M33 \\
& no DM exc & no DM exc - STH2 & no DM exc - NFW &  no DM exc - NFW \\
\hline
$[0,1]\ R_{\rm vir}$ & 2.3 (1.1) & 1.6 (1.0) & 1.6 (0.7) & 0.5 (0.0) \\
$[1,1.5]\ R_{\rm vir}$ & 1.7 (4.1) & 1.2 (3.2) & 1.2 (3.0) & 0.8 (2.3) \\
$[1.5,2]\ R_{\rm vir}$ & 1.3 (0.8) & 0.8 (0.8) & 0.9 (0.8) & 0.7 (0.5) \\
total & 6.3 (6.0) & 4.5 (5.0) & 4.6 (4.5) & 2.6 (2.8) \\
\hline
\hline
$10^{12}-10^{13}\ M_\odot$ & $\chi^2$ & $\Delta\chi^2$ & $\Delta\chi^2$ & $\Delta\chi^2$ no M31 \\
& no DM exc & no DM exc - STH2 & no DM exc - NFW &  no DM exc - NFW \\
\hline
$[0,1]\ R_{\rm vir}$ & 1.7 (2.0) & 1.5 (1.9) & 1.4 (1.9) & 1.6 (1.9) \\
$[1,1.5]\ R_{\rm vir}$ & 0.3 (0.0) & -0.7 (-1.3) & -0.5 (-1.0) & -0.6 (-0.5) \\
$[1.5,2]\ R_{\rm vir}$ & 1.0 (1.2) & 0.8 (1.2) & 0.8 (1.2) & -1.1 (-1.2) \\
total & 3.1 (3.2) & 1.8 (1.8) & 1.7 (2.2) & -0.3 (0.1) \\
\hline
\end{tabular}
\end{table*}

Table~\ref{tab:chi2} lists the $\chi^2$ values as well as their differences, $\Delta\chi^2$, in each radial bin.  It also lists the total $\chi^2$ summing over all bins.  By default, we calculate $\chi^2$ by marginalizing localization errors, but we show the $\chi^2$ computed using maximum likelihood localization in brackets (i.e. assuming $P(\mathbf{b})$ is a $\delta$-function at the best fit value).\footnote{We note that technically $\chi^2 \geq \sum_j \chi^2_j$, with equality achieved when $P(\mathbf{b})$ is a delta function.  This is why the total $\chi^2$ differences are larger than summing up the $\Delta\chi^2$ values in individual bins.}  Not marginalizing over localization errors generally changes the results at the $<1\sigma$ level, but in some cases the differences are somewhat larger showing that accounting for positional errors is important.  The $\Delta\chi^2$ values are calculated by subtracting the $\chi^2$ of the no DM excess model with that of the STH2 or NFW model.  We have also verified that further dividing the radial bins (but keeping the number of FRBs in a bin at least 15-20) do not affect the $\Delta\chi^2$ values.
With 3 radial bins we would expect total $\chi^2\sim3$, and the models produce a slight over-fit of total $\chi^2\sim1.5$.

The $\chi^2$ in all radial bins of the $10^{11}-10^{12}\ M_\odot$ halos disfavors the no DM excess model at $\sqrt{\Delta\chi^2}\approx1$-$\sigma$ level, although the significance level drops when excluding M33.  The total $\chi^2$ of NFW and STH2 models considering all bins is smaller than that of the no DM excess case by 2-5, regardless of marginalizing over the localization uncertainties or using maximum likelihood localization.  Thus the significance that a DM excess is preferred over no excess ranges from 1.4-2.2\,$\sigma$.  The $10^{12}-10^{13}\ M_\odot$ halos, on the other hand, favor a DM excess only in the $[0, 1]\ R_{\rm vir}$ bin and cannot distinguish any models for the larger impact-parameter bins.  The $\Delta\chi^2$ values are consistent with our results in Figure~\ref{fig:observed_hists_and_vals}. 
Finally, the $\chi^2$ of the STH2 and NFW models are very similar, as anticipated, and so not distinguishable by the data.

\subsection{How many FRBs are needed to get a significant detection of DM excess}
\label{sec:prediction}

Thus, we have found a marginal detection of an excess DM when stacking on foreground halos, but the data set of 453 FRBs is insufficient to distinguish between realistic CGM models.  Larger samples are of course required to get a more significant detection of a given DM excess of 50-100 -- the anticipated DM excess for the CGM of $10^{11}-10^{13}\ M_\odot$  halos and this is also the difference between viable models for $b<0.5\ R_{\rm vir}$.
Since we currently get an error bar in DM of 50 with $\approx 30$ intersections, we anticipate that the error bar scales with the number of intersections as $\sigma_{\rm DM}= 50\, \sqrt{30/N_{\rm DM}}$.  With 200 intersections then $\sigma_{\rm DM}=20$; this many intersections would require about 3000 FRBs in the next CHIME data release.  

Currently the number of sightlines passing through halos at $<0.5\ R_{\rm vir}$ is too small, but this is the region where viable models for the CGM gas profile differ the most.  With about $10^4$ more FRBs from CHIME, the number of intersections with $10^{11}-10^{13}\ M_\odot$ halos at $b<0.5\ R_{\rm vir}$ will reach $\sim200$.  Such a sample would be able to put novel constraints on CGM models.



\section{Conclusions}
\label{sec:conclusions}

We have measured the DM excess owing to the CGM of $10^{11}-10^{13}\ M_\odot$ halos at $<80$~Mpc using the CHIME/FRB first data release.  To this end, we have developed a weighted stacking scheme to reduce the variance of the observed DM distribution and to lower the bias of high-DM FRBs on the sample mean DM.  With 20-30 FRBs intersecting $10^{11}-10^{12}$ and $10^{12}-10^{13}\ M_\odot$ halos at $<1\ R_{\rm vir}$, we find that the DM excess of these halo groups can be detected at 1-2\,$\sigma$.  We also tentatively detect a DM excess at impact parameters of 1-2\,$R_{\rm vir}$ for $10^{11}-10^{12}\ M_\odot$ halos, but not for the $10^{12}-10^{13}\ M_\odot$ halos.  
With more FRB data from CHIME coming in the near future, each stack's signal-to-noise ratio will continue to improve with an error of $\sigma_{\rm DM}= 50\, \sqrt{30/N_{\rm DM}}$~pc~cm$^{-3}$, where $N_{\rm DM}$ is the number of intersections (which was approximately $30$ in our stacks).

We have also calculated the likelihood of different CGM models given the CHIME FRB data.  All models are favored by the data at 1.4-2.2\,$\sigma$ over a model with no DM excess, consistent with our measurements of a DM excess.  We find that viable models for the CGM gas distribution produce DM differences smaller than 50 outside $0.5\ R_{\rm vir}$ because of the importance of the two-halo term around $10^{11}-10^{13}\ M_\odot$ halos.  Owing to the paucity of intersections at these impact parameters, a large increase in the number of FRBs to at least ten thousand would be needed to discriminate between viable CGM model with our method that stacks on nearby galaxies.  

In the next few years, our weighted stacking method will continue to be a valuable tool to detect and measure the CGM of nearby halos as more data from CHIME comes along.  Especially for the relatively low-mass ones as we considered here, whose CGM is hard to probe in any other way.  With new surveys launching that aim to observe FRBs with arcsecond localizations, other methodologies may provide better constraints on the CGM gas profile.  For instance, the Canadian Hydrogen Observatory and Radio-transient Detector \citep[CHORD][]{chordwhitepaper} will receive $>20$ bursts per day, with the goal to provide milli-arcsecond localization accuracy of FRBs with VLBI.  DSA-2000 will also start observing $\sim75\%$ of the full sky with arcsecond spatial resolution \citep{dsa2000whitepaper}.  Once localized to a galaxy, the mean cosmic dispersion to a redshift can be removed from each FRB, which will dramatically reduce the noise in the stack allowing constraints with just hundreds of bursts \citep{mcquinn14,williams22}.  It might even be possible to forward model each component (intergalactic, Milky Way, host galaxy) of the FRB DM instead of performing weighted stacking.  Especially, the intergalactic contribution might be modeled with a reconstruction of the underlying density field \citep[e.g.][]{burchett20}.  The improved angular resolution of these surveys will also allow stacking on galaxies that are further away than in this study.  Another interesting related direction is to constrain the fraction of cool ionized gas via scattering and lensing of FRBs \citep{vedantham19, prochaska19}. 

\section*{Acknowledgements}

We thank Ian Williams for providing the CGMBrush profiles, Matt Wilde for offering the NSA galaxy catalog, and Bryan Gaensler, Yakov Faerman, Ue-Li Pen, Sandro Tacchella for useful discussions.
We acknowledge support from NSF award AST-2007012.

\section*{Data Availability}

The data underlying this article will be shared on reasonable request to the corresponding author.



\bibliographystyle{mnras}
\bibliography{references} 




\appendix

\section{Possibility of FRBs originating from nearby halos}
\label{sec:discussion}

While we have implicitly assumed that all the FRBs used in this work should originate from halos further away than 40-80 Mpc, it is likely that some FRBs may come from nearby halos, especially low-DM (DM~$\lesssim100$) ones \citep{bhardwaj21a, bhardwaj21b}.  Even for FRBs with DM of several hundred, it is not unlikely that they are from nearby galaxies because the host DM may well be a few hundred \citep[Table 2 of][]{cordes22}.

To address this issue, we ran the Probabilistic Association of Transients to Hosts (PATH) code\footnote{\url{https://github.com/FRBs/astropath}} to determine which FRBs might originate from $<80$ Mpc galaxies \citep{aggarwal21}.  PATH calculates the probability that an extragalactic transient source is associated with a candidate host galaxy using the Bayes' rule.  For each FRB, we take all $<80$~Mpc galaxies in GWGC that lie within 3 times the FRB RA Dec error-bars to be the candidates which the FRB can be associated with.  Instead of using the ``inverse'' prior that assumes brighter galaxies have higher probabilities, we adopt a uniform prior.  We assume that the distribution of transients around galaxies follows an exponential profile, where the size of the exponential function is given by the tabulated major and minor diameters of galaxies in GWGC.  PATH then integrates over the FRB localization ellipse, and finds the galaxy with a $>0.95$ posterior to be the most likely host of the FRB.  It does not take into account the FRB DM, however.

For the 26 (30) FRBs that we find to pass through nearby $10^{11}-10^{12}\ M_\odot (10^{12}-10^{13}\ M_\odot)$ halos at $<1\ R_{\rm vir}$, PATH shows that 6 (1) of them might originate from these halos themselves.  We find 7 more FRBs likely associated with the intersected galaxies when focusing on $b<2\ R_{\rm vir}$, and 3 FRBs may arise from halos closer to the ones they are found to intersect with.  We thus performed our measurement of DM excess again without these FRBs identified, and find that our main conclusions remain unchanged.  The PATH code, however, is only a rough way of estimating the probability of FRBs being associated with galaxies and much more careful visual inspection on the locations of galaxies versus the localization contours of FRBs should be done before drawing conclusions.


\bsp	
\label{lastpage}
\end{document}